# All downhill from the PhD? The typical impact trajectory of US academic careers[1]

Mike Thelwall, Ruth Fairclough, University of Wolverhampton, UK.

Within academia, mature researchers tend to be more senior, but do they also tend to write higher impact articles? This article assesses long-term publishing (16+ years) United States (US) researchers, contrasting them with shorter-term publishing researchers (1, 6 or 10 years). A long-term US researcher is operationalised as having a first Scopus-indexed journal article in exactly 2001 and one in 2016-2019, with US main affiliations in their first and last articles. Researchers publishing in large teams (11+ authors) were excluded. The average field and year normalised citation impact of long- and shorter-term US researchers' journal articles decreases over time relative to the national average, with especially large falls to the last articles published that may be at least partly due to a decline in self-citations. In many cases researchers start by publishing above US average citation impact research and end by publishing below US average citation impact research. Thus, research managers should not assume that senior researchers will usually write the highest impact papers.

## 1 Introduction

US university departments seem to be managed by, and partly populated by, experienced tenured researchers, who, if they continue publishing, also have greater access to resources (e.g., Levitt & Levitt, 2017). In this context, and in conjunction with gaining knowledge through long term participation in a field, it would be reasonable to expect that experienced US academics would produce higher citation impact research. Knowledge about the extent to which this is true would be useful for research managers deciding on the optimal balance of junior and senior researchers or the types of activities that would make the best use of senior researchers' time.

Previous studies of academic careers have tended to be qualitative or cover individual fields, investigating different factors and producing divergent findings (Sugimoto, Sugimoto, Tsou, Milojević, & Larivière, 2016). For junior researchers, effective mentoring has been shown to be helpful for long term publishing prospects in neuroscience and biomedical science (Liénard, Achakulvisut, Acuna, & David, 2018) and the same is true for early collaborations with highly cited scientists in four disciplines (Li, Aste, Caccioli, & Livan, 2019). A comparison of 100 junior and 200 senior physicists suggested that bad luck might prematurely terminate careers (Petersen, Riccaboni, Stanley, & Pammolli, 2012). A survey of 624 US plastic surgeons found that those publishing more during their training also published more afterwards (DeLong, Hughes, Tandon, Choi, & Zenn, 2014), with the same found for medicine, science and technology at one Swedish university (Lindahl, Colliander, & Danell, 2020). Early research grants moderately associate with higher career impact for three social science fields in the Netherlands (Van den Besselaar & Sandström, 2015).

It seems likely that there are substantial overall differences in the trajectories of careers, associated with, for example, international differences in the financial support for

---

and growth-rate of higher education (Finkelstein, 2015), as well as disciplinary differences in publishing rates (Larivière, & Costas, 2016), output types (Verleysen & Ossenblok, 2017), collaboration types (Lewis, Ross, & Holden, 2012), research group culture (Pull, Pferdmenges, & Backes-Gellner, 2016), mentoring (Ooms, Werker, & Hopp, 2019), and gender (Fox & Stephan, 2001; Thelwall, Bailey, Tobin, & Bradshaw, 2019). Finally, senior transportation researchers produce higher citation impact papers (Hanssen & Jørgensen, 2015).

Highly cited researchers have been subjected to special attention for career factors. Analyses of 450 highly cited scientists suggested that their reputation boosted the citation rates of their later papers (Petersen, Fortunato, Pan, Kaski, Penner, Rungi, & Pammolli, 2014; see also: Petersen, Jung, Yang, & Stanley, 2011), and a logical consequence of this is that average citation rates would increase with academic age, for highly cited researchers. Nevertheless, an investigation of highly cited scientists in seven disciplines did not find a temporal pattern in the citation impact of their work (Sinatra, Wang, Deville, Song, & Barabási, 2016).

A range of career influences on citation impact have also been investigated for other sets of researchers. Sociology, politics or political science elite US institution faculty CVs (n=1002) have been used to analyse academic careers, finding that productivity (average number of journal articles) was relatively constant, the proportion of collaboratively authored articles increased and average (arithmetic mean) citation rates decreased over time (not statistically significant) (Sugimoto, Sugimoto, Tsou, Milojević, & Larivière, 2016). Nevertheless, an analysis of 6,388 Quebec professors found that the average (arithmetic mean) field normalised citation impact of their papers was lowest at age 50, when their productivity (articles per year) also peaked (Gingras, Lariviere, Macaluso, & Robitaille, 2008). Similarly, for Spanish research council scholars, those scoring highest on a composite indicator based on a range of productivity and citation indicators tended to be younger than average (Costas, van Leeuwen, & Bordons, 2010). In contrast, although with different methods, in Mexico total citations over a four-year period peaked at age 56 for researchers (González-Brambila & Veloso, 2007).

Second order factors, such as collaboration, gender, mobility or productivity may influence citation rates differently over time. The most researched age-related factor seems to be productivity rather than citations, however (e.g., Kyvik & Olsen, 2008; Mishra & Smyth, 2013; Rørstad & Aksnes, 2015). The number of collaborators of a computer scientist or physicist increases with academic age (Wang, Yu, Bekele, Kong, & Xia, 2017), and collaboration overall in academia associates with more citations (Larivière, Gingras, Sugimoto, & Tsou, 2015). Based on 375 US researchers, senior chemists seemed to maintain high productivity at the expense of reduced average citation impact, but in mechanical engineering higher productivity associated with higher average research impact (Kolesnikov, Fukumoto, & Bozeman, 2018). In Japan, older researchers write less papers, perhaps because they have less time for research (Kawaguchi, Kondo, & Saito, 2016), but it is not known if this influences their average citation impact. In the USA, female researchers can expect to attract marginally more citations for their research (Thelwall, 2018b). The average citation rate for physicists moving to less prestigious institutions was found to decrease in another study, but the reverse was not true (Deville, Wang, Sinatra, Song, Blondel, & Barabási, 2014).

Productivity is a key second-order factor for citation impact. As mentioned above, there is conflicting evidence about whether highly cited researchers may attract a late career citation boost (Petersen, et al., 2014; Sinatra, et al., 2016). Nevertheless, the Matthew effect suggests that "established researchers" have an advantage when attracting funding, citations and credit (Merton, 1968). They would presumably also tend to publish more, due to increased resources, opportunities for collaboration and motivation. If such researchers are more likely to continue publishing than others, then the citation impact and productivity of longer-term researchers may tend to be higher. It is not clear how this would affect career trajectories for citations, since a researcher gaining from the Matthew effect is likely to have published highly cited early work to become established and their "extra" Matthew effect citations could target their early or later work. If researchers can become established through high productivity rather than high quality work (Merton, 1988), then the Matthew effect would create an additional connection between productivity and average citation rates. For 48,000 author disambiguated Swedish researchers based on four publication years (2008-11) in the Web of Science, higher productivity associates with a higher probability for each publication to be more cited for its field and year (more specifically, to fall in a higher field normalised citation class: Sandström & van den Besselaar, 2016), supporting a positive association between productivity and citation impact. This relationship was also found for an international set of 28 million Web of Science (WoS) authors over 34 years (1980–2013): more publications per year associated with a higher probability for each one to be in the top 1% cited for its field and year in the Medical and Life Sciences, Social and Behavioural Sciences, and Natural Sciences but not Law, Arts and Humanities (Larivière & Costas, 2016).

Some studies have analysed complete or complete to date academic publishing careers at the national level. For three universities in Australia, on average, researchers at the end of their careers publish in higher impact journals than researchers at the start of their careers (Fig 18 of: Gu & Blackmore, 2017), but it is not clear whether this is because less successful early career researchers leave academia or stop publishing. The average (arithmetic mean) citation impact is highest mid-career (Table 4 of: Gu & Blackmore, 2017), but the results are not field normalised. There is a relatively small difference in the likelihood of being a first author between career stages, although junior and senior researchers are more likely to be first authors than mid-career researchers (Fig 16 of: Gu & Blackmore, 2017). Combining this with the results reported in previous paragraphs (younger researchers more cited in Spain and Quebec, older researchers more cited in Mexico, all using different methods and samples), there is not a clear relationship between average citation impact and either (academic or physical) age or career stage. The above-mentioned international WoS study found that the likelihood of an article being in the top 1% cited for its WoS field was higher for authors with longer publishing careers (Larivière & Costas, 2016). This could be a second-order effect of international differences, however, for example if US researchers had longer careers, and published higher impact work.

This article assesses changes over time in the average citation impact of US researchers based on their publishing career duration. Based on the above discussion it is not known whether, in general, research impact per paper tends to change during (publishing) careers and whether any change is influenced by the duration of those careers. The focus is on the USA for methodological pragmatism: the most robust career publishing data is available for this country, as described in the methods.

- RQ1: For long-term US researchers, how does the average citation impact of their work vary during their careers?
- RQ2: Does the answer to the above question change for shorter-term researchers?

## 2 Methods

The research design was to identify separate sets of long-term US researchers at a given start date through their Scopus IDs and publication record in Scopus, identifying any citation impact changes over time. The USA is a suitable initial case study because it is a large country with a research-intensive culture (hence a large amount of data) and international citation indexes seem to have relatively extensive coverage of English-language publications and US journal publishers (Mongeon & Paul-Hus, 2016); coverage for researchers in advanced non-English speaking countries seems likely to be lower due to language issues (e.g., Kulczycki, Guns, Pölönen, Engels, et al., in press). In addition, as a rich nation, successful researchers seem less likely to leave to a better funded position elsewhere. For example, successful long-term researchers from poorer nations might be tempted to move to the US for higher salaries or better research support (although researchers can collaborate internationally instead, e.g., Kwiek, in press). Less developed countries are, in general, less attractive targets for researcher mobility (IDEA Consult, WIFO, & Technopolis, 2017), with the USA being particularly attractive place to research (Janger, Strauss, & Campbell, 2013). One study of 12,502 full professors at elite US institutions found that 88% had US PhDs, although 22% had moved to the US after their undergraduate degree (Yuret, 2018). This occurs despite extensive geographic mobility within the US (He, Zhen, & Wu, 2019).

Scopus was chosen in preference to the Web of Science for more comprehensive coverage (Mongeon & Paul-Hus, 2016) and in preference to Dimensions (Thelwall, 2018a), Google Scholar (Harzing & Alakangas, 2016) and Microsoft Academic (Harzing & Alakangas, 2017) due to having a standard classification scheme that is known to be reasonably consistent (Klavans & Boyack, 2017) and substantial long term coverage. Scopus records from 1996 to 2013 were downloaded in late 2018, and Scopus records from 2014 to 2019 were downloaded in early 2020 so every citation count used is based on at least three full years of citations. An open citation window was used to give the maximum statistical power to the results. Only documents of type Journal Article in Scopus were included since these are the primary research outputs in most fields. The exclusion of review articles is a limitation since these can make important contributions (Yadav, 2010), but it does not seem possible to accurately field normalise their citation counts due to their relative scarcity. Similarly, the omission of conference papers, monographs and edited books are limitations because they are central to some fields but there do not seem to be scholarly databases that can be connected to Scopus (in terms of author IDs) with a substantial fraction of academic books.

Researcher track records were traced from 1996, after an expansion of Scopus, and a researcher was assumed to have started their career in 2001 if they had not published any co-authored articles in Scopus 1996-2000 but published at least one in 2001. This is an oversimplification because a researcher may have had a period of inactivity or may have produced other forms of outputs 1996-2000. Nevertheless, it seems reasonable as a way of selecting researchers that are likely to have started publishing in 2001. The number of years since a first publication is sometimes called academic age (e.g., Milojević, 2012).

Scopus researcher IDs were used to track individual researchers throughout the period. These are imperfect but seem to be highly accurate overall (Aman, 2018; Kawashima & Tomizawa, 2015; Strotmann & Zhao, 2012). Intuitively, they may be least reliable for the start of a career, when researchers can move between institutions after their PhD, between postdoctoral positions, and to a permanent post. Thus, it is possible that long-term researchers (defined as below) sometimes have Scopus track records omitting their earliest work. Similarly, some short-term researchers in Scopus (defined as below) may instead represent the career starts of longer-term researchers. Each publication listed with an author's ID in Scopus was assigned to that author. The assignment process is imperfect and seems to be limited to 100 authors per paper, but seems likely to be accurate for almost all papers, except perhaps in astronomy and high energy physics, where collaborations with hundreds or thousands of authors are common. This problem has been ignored so the results may include researchers that should have been excluded for participating in large team collaborative research. This seems unlikely to make much difference to the results, because very large team papers seem to be rare and may represent a different type of activity for researchers that also produce smaller team research.

A long-term researcher was defined to be one with a first Scopus-indexed publication in 2001 (to give a common start year, and prior publications checked back to 1996) and at least one Scopus-indexed publication in 2016-19, to ensure that they were active for at least 16 years. A researcher was classed as from the US if the affiliation of their first publication and last publication was within the USA. This includes non-US researchers that moved to the USA for a PhD and then stayed there (or left and returned). The US affiliation could be for any type of organisation, rather than just universities. This seems to be the most relevant group from a US policy perspective. Researchers were excluded if any of their publications had more than ten authors. This step was designed to exclude highly co-authored research for which co-authorship may not be a good indicator of contribution. This affects highly collaborative areas of genomics, astrophysics, and high energy physics for which co-authorship seems to be procedural. Researchers were also excluded if they had published less than five articles since these would be relatively inactive from a publishing perspective and would therefore provide little mid-career publishing evidence.

To test for the accuracy of the above method to identify each researcher's first publication year, a random sample of 100 of the matching long-term US researchers (selected with a random number generator) were searched for in Scopus by ID and their earliest publication date checked. In 21 cases there was a Scopus publication before 2001, with the earliest being 1967. Thus, about a fifth of the sample of long-term researchers have an earlier initial publication date than found by the method used here.

Medium- and short-term researchers were defined as above, except that the last publication had to be on a specified date so that the exact duration of their publishing career was known. Medium term was defined to be 10 publishing years and short term 6 publishing years. In both cases multiple starting years are allowed, beginning with 2001 and ending not after 2016, giving multiple cohorts. The years 6 and 10 were chosen as illustrative intermediate values between 1 and 19 rather than for any theoretical reason. Researchers publishing a single journal article were also analysed as very short-term researchers (Table 1). The team size restriction (excluding researchers ever authoring in teams of 11+) excluded most of the 15,329 long term researchers, giving a final data set size of 5825. Of these, only 771 had their most recent journal article published in 2016 (the remainder had articles published in 2017, 2018 or 2019).

Table 1. Cohort sizes for the main and supplementary datasets analysed. Unless specified, researchers are excluded if any of their publications have 11+ authors.

| Group | Average researchers per cohort | Min | Max |
| --- | --- | --- | --- |
| Long term researchers | 5825 | 5825 | 5825 |
| Medium term researchers | 814 | 585 | 1155 |
| Short term researchers | 1042 | 711 | 1691 |
| Single paper researchers | 70204 | 51524 | 94980 |

The average citation impact of the publications produced by the chosen researchers in each field and year was calculated using the Mean Normalised Log Citation Score (MNLCS) (Thelwall, 2017) variant of the Mean Normalised Citation Score (MNCS) (Waltman, van Eck, van Leeuwen, Visser, & van Raan, 2011). The MNLCS calculation log transforms all citation counts with $ln(1+c)$ because citation data is skewed, and the results would otherwise reflect highly cited papers to some extent rather than average behaviour. The formula divides each logged citation count with the world average logged citation count for the field and year of publication (or the average of the world averages for papers assigned to multiple fields). Scopus narrow fields were used for the normalisation, with 330 separate fields in each year, giving fine-grained field normalisation. Researchers producing more than one paper in a given year contributed the average impact of these papers, rather than each paper separately. Researchers not producing any papers in a year were ignored for that year. Confidence intervals were calculated for MNLCS values using the normal distribution formula, which is appropriate because the log transformation greatly reduces the skewing in the citation count data. The formula is likely to be a bit conservative because some of the data points are averages over multiple papers by the same researcher, reducing variation. Articles can be counted multiple times if they have multiple qualifying authors, since the focus is on the average per researcher rather than per author. The resulting MNLCS is above 1 only if the research has more impact than the world average and can fairly be compared between years.

## 3 Results

The main results are introduced first, with factors potentially influencing them discussed afterwards.

### 3.1 Citation impact trends over careers

Long term US researchers (16+ years of journal article publishing) begin with citation impact above the US (and world) average but the citation impact of their articles declines steadily until it is below average for the USA (but above world average) (Figure 1). Thus, whilst long term US researchers tend to be able to produce high citation impact research, this capability apparently declines with age unless other factors explain the decrease. The USA average here is the MNLCS for all publications from US researchers (defined as having a first and last publication with a US affiliation), irrespective of first and last publication date, except excluding researchers that have ever co-authored in teams of 11 or more. The USA average reference set is therefore calculated from the publications of comparable researchers.

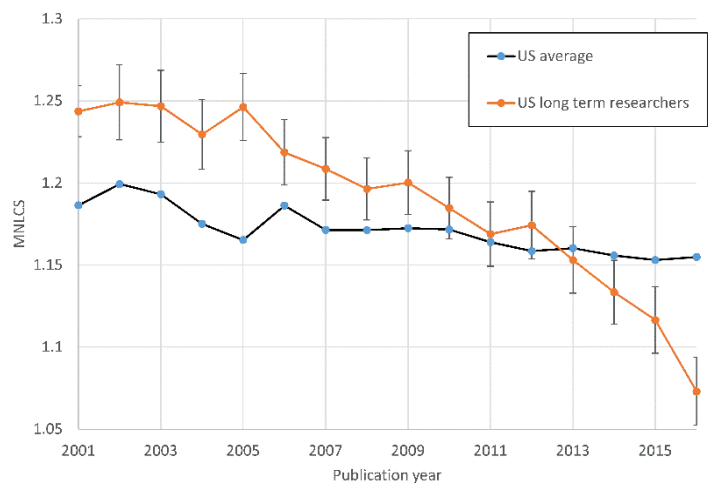

Figure 1. Average citation impact of publications co-authored by US long term researchers (first Scopus publication in 2001; at least one publication 2016-19; first and last publication with a US affiliation, at least 5 publications, ignoring researchers ever co-authoring in teams of 11+; n=5825 researchers). Error bars show 95% confidence intervals. The reference set is researchers with first and last publication with a US affiliation, ignoring researchers ever co-authoring in teams of 11+ (n=643,204 researchers).

As with long-term researchers, medium-term researchers (exactly 10 years of journal article publishing, according to Scopus) tend to initially publish above average citation impact journal articles but their last articles have substantially lower citation impact on average (Figure 2). The pattern is broadly similar irrespective of starting year. Confidence intervals are not shown (available in the supplementary material) but are wider than for Figure 1, so the occasional increases in the line heights could be statistical anomalies.

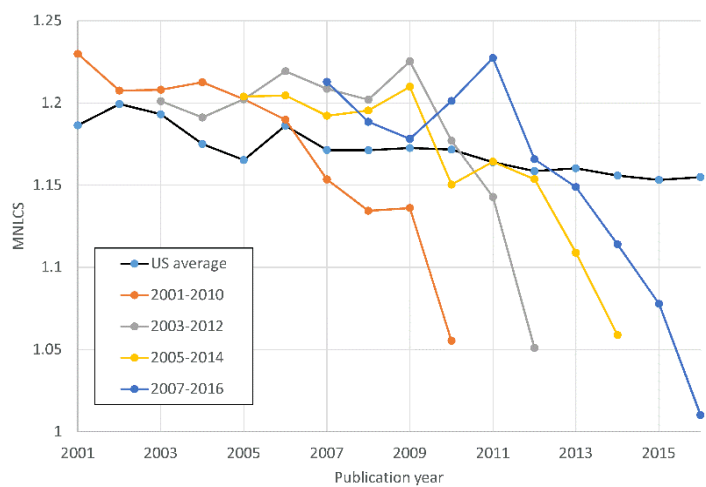

Figure 2. Average citation impact of publications co-authored by four cohorts (585 to 1155 researchers per cohort) of US medium term (10 years of Scopus journal article publishing) researchers (first and last Scopus publication as specified in legend; first and last publication with a US affiliation, at least 5 publications, ignoring researchers ever co-authoring in teams of 11+).

Short-term researchers (exactly 6 years of journal article publishing, according to Scopus) tend to initially publish above average citation impact journal articles but with average citation impact decreasing over time (Figure 3). For consistency, the five-publication minimum requirement applies to Figure 3 and Figure 2, so the short term researchers in Figure 3 have a higher productivity requirement and may therefore be more capable, on average, than short term researchers without this requirement.

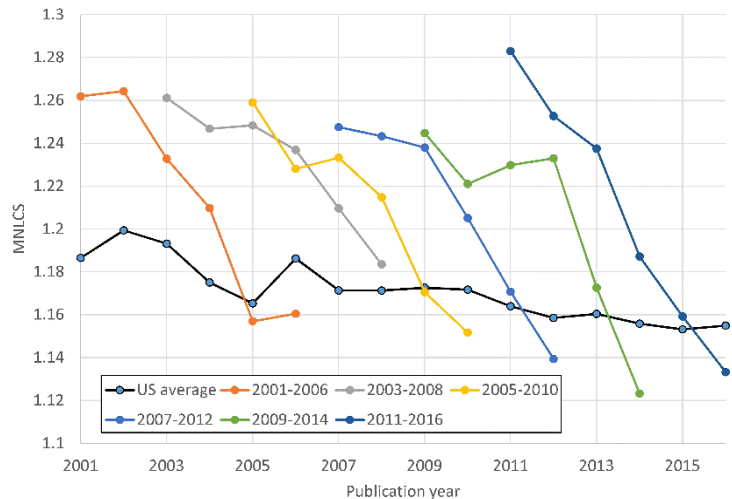

Figure 3. As Figure 2 for six cohorts of short term researchers (6 years of Scopus journal article publishing; 711 to 1691 researchers per cohort).

US authors publishing only a single article indexed in Scopus tend to produce work that has a low citation impact for the USA (Figure 4). Thus, the early career high citation impact for long, medium and short-term researchers does not apply in this case.

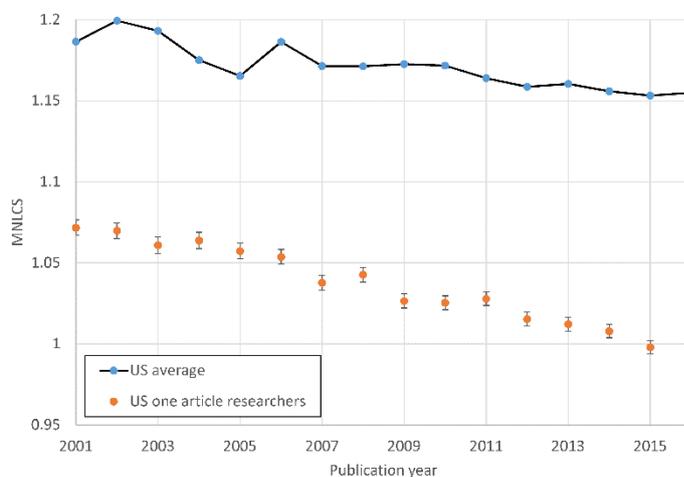

Figure 4. Average citation impact of publications co-authored by US researchers only ever publishing a single article in Scopus (ignoring researchers co-authoring in teams of 11+).

### 3.2 Factors associating with citation impact

Team size influences citation impact. If the restriction on the maximum team size for researchers is removed then the average impact of long-term researchers is much higher (Figure 5, compare with Figure 1) but still has a consistent downward trend in average citation impact after 2009, especially compared to the new reference set without a team size restriction (which also has a higher average citation impact) in Figure 5. Recall that this is a less reasonable graph because it includes researchers that have only made contributions to large team studies, perhaps occasionally in a minor role. The MNLCS dip in the reference set for 2014 and 2015 is presumably due to a change in the journals indexed in Scopus that particularly influences highly collaborative publications.

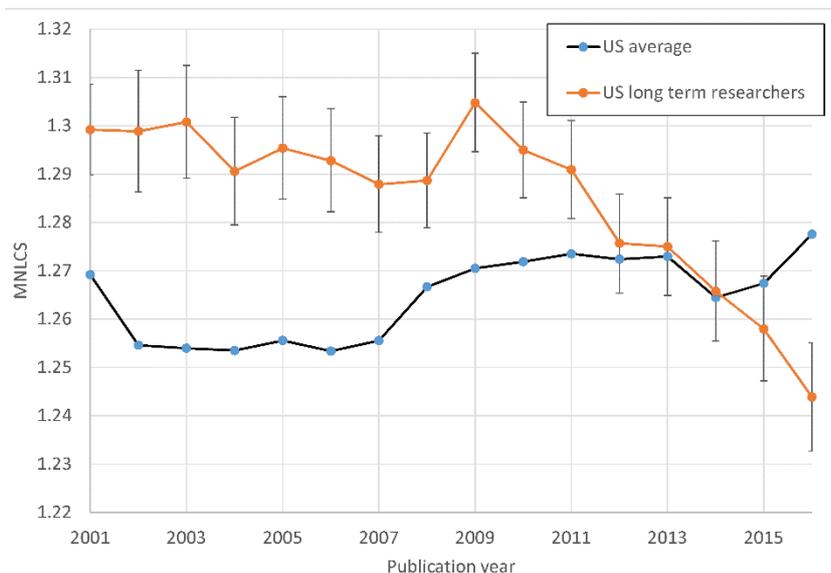

Figure 5. As Figure 1, but without any team size restriction. The reference set is as in Figure 1 but ignoring the team size restriction.

If the restriction on the minimum number of papers is removed (i.e., reduced from 5 to 2, to accommodate the initial and final publications) then this has no effect on the results (Figure 6).

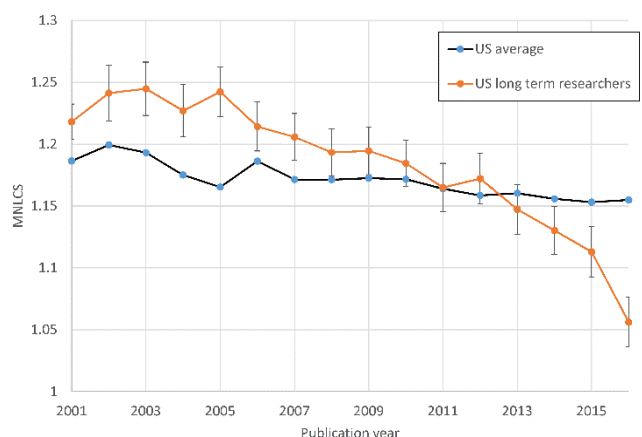

Figure 6. As Figure 1, but without any productivity restriction (minimum 2 publications, one in 2001, one on or after 2016).

The extent to which a long-term researcher collaborates may vary during their career. Any such changes could explain changes in average citation impact since more collaborative articles tend to be more cited (Larivière, Gingras, Sugimoto, & Tsou, 2015). The overall trend is for the rate of collaboration for long term researchers to increase over time at a slightly higher rate than the US average (Figure 7), which does not explain the decreasing impact in Figure 1. It is perhaps surprising that long term researchers collaborate less than average researchers in the US (a lower line in Figure 7). This might be due to non-researchers being occasionally added to larger collaboration teams for specialist services (e.g., medical doctors advising a survey; a technician making a particularly useful piece of equipment). Another possibility (suggested by a reviewer of this paper) is that early career researchers may need to collaborate initially as they learn but then leave academia if they fail to develop as independent researchers.

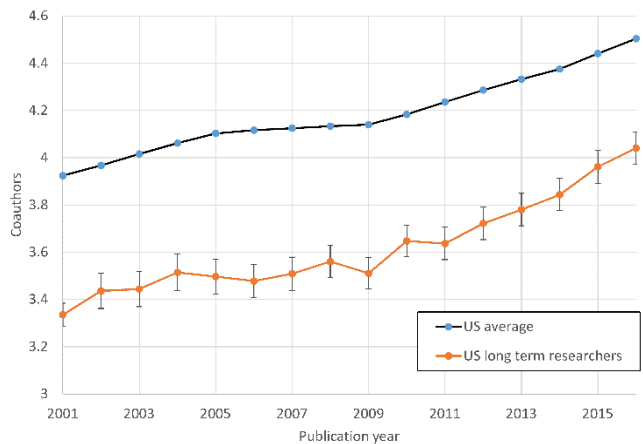

Figure 7. As Figure 1, but for the average (geometric mean) number of authors per paper.

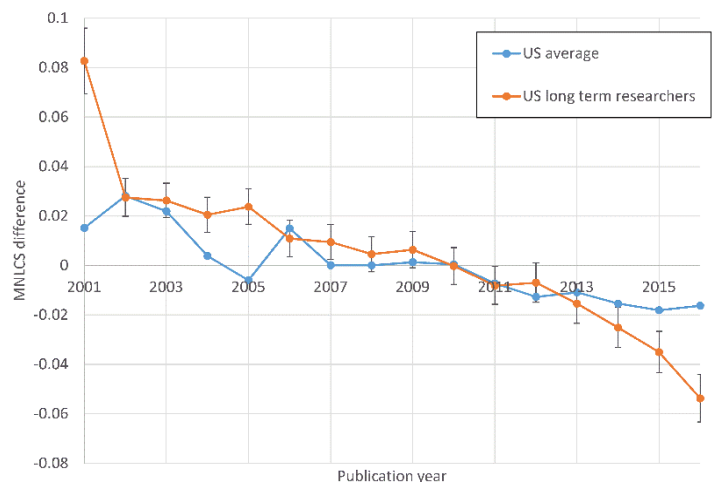

Figure 8. Average citation impact of publications co-authored by US long term researchers subtract each researcher's average over the period (first Scopus publication in 2001; at least one publication 2016-19; first and last publication with a US affiliation, at least 5 publications, ignoring researchers ever co-authoring in teams of 11+). Error bars show 95% confidence intervals. The reference set is researchers with first and last publication with a US affiliation, ignoring researchers ever co-authoring in teams of 11+, with the average MNLCS subtracted from each year.

For medium-term researchers, after factoring out productivity as described above, there is a decrease in average impact for all cohorts between the first and subsequent publication years (Figure 9) and the same applies for short term researchers (Figure 10). Thus, impact drops after the first publication year and at the end of publishing careers seem to be universal for researchers with at least short-term careers.

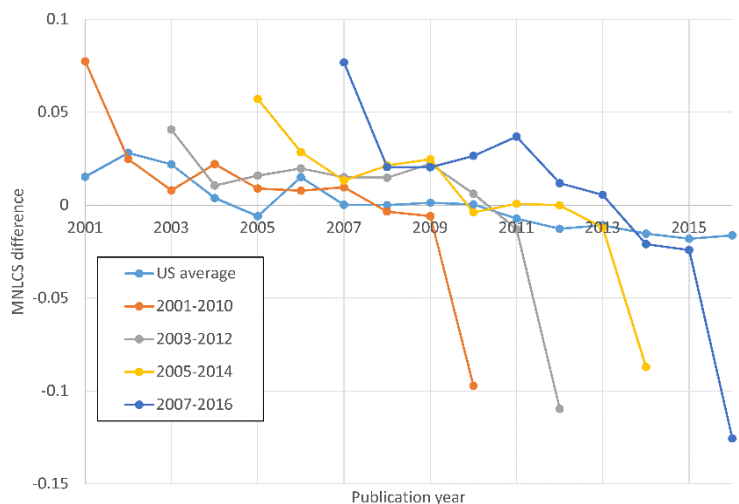

Figure 9. Average citation impact of publications co-authored by four cohorts of US medium term researchers, subtract each researcher's average over the period (first and last Scopus publication as specified in legend; first and last publication with a US affiliation, at least 5 publications, ignoring researchers ever co-authoring in teams of 11+).

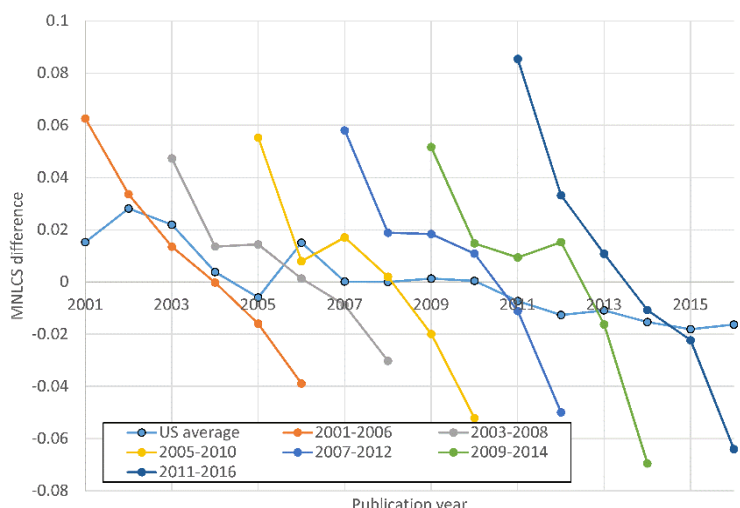

Figure 10. As Figure 9 for six cohorts of short term researchers.

Confidence intervals for all figures as well as data for additional figures are available in the online supplementary materials https://doi.org/10.6084/m9.figshare.11791059.

## 4 Discussion

The results are limited by the data source (Scopus), the accurate but imperfect author identification procedure used by Scopus, the limitation to Scopus-indexed journal articles, the analysis of articles using whole author counting, the restriction to authors that have never published in large teams (11+ authors), and the aggregate cross-discipline reporting. For Figure 1 and all results with data starting from 2001, since 21% of authors had earlier publications than found by the method here (before 1996), the trends may be stronger than shown in the graphs. This is because the first year of publication tends to have the highest citation impact and so including publications from authors not in their first publishing year would tend to reduce the average impact of the set. It would add to the robustness of the results if they could be replicated with a citation database that is at least as large, differentiates different publication types, and includes author IDs or enough information to systematically disambiguate author names, but no other database satisfies these criteria. Lacking this and taking into account the exclusions of reviews, conference papers and academic books weakens the strength of the evidence. Nevertheless, for the trends identified to be incorrect, there would need to be systematic academic age-related factors related to these issues that run strongly counter to the results reported above, which seems unlikely. The trends for long term researchers may also vary between cohorts, which has not been tested for. There are also likely to be disciplinary differences in the trends, which have also not been tested for. The citation impact trends should not be used to judge the wider value of the work of the researchers. This is because citations only reflect one type of research impact and academics can make contributions in ways other than publishing journal articles, such as through education, mentoring and management. For the shorter career paths, trained researchers can also make valuable contributions to society outside of the academic publishing model. The findings should also not be extrapolated beyond the USA. There are many reasons why these findings are unlikely to replicate to all countries. On an international scale, the US system seems to be relatively competitive (Angermuller, 2017), as well as being a large, rich nation with long research track record and relatively comprehensive journal coverage in scholarly databases.

The results show that the average field normalised citation impact of US researchers declines over the course of their career, and particularly towards the end. After factoring out the productivity effect, there is also a sharp impact drop after the first publishing year. This applies to long, medium and short-term researchers. It is not affected by the restrictions on team size and productivity chosen to make the default analysis more useful (results are per paper, not per academic, so publishing multiple papers in a year is not an advantage for the figures reported). It is not a side-effect of collaboration, since collaboration increases over career length, and is known to associate with higher citation impact (Larivière, et al., 2015). The overall decline agrees with a study of three social sciences in the US (Sugimoto, et al., 2016) but sharply contrasts with the increase in field normalised average citations after age 50 found in Quebec, Canada (Gingras et al., 2008) either because of country/language issues or the use of the arithmetic mean in the latter study. The overall decline also contrasts with the mid-career average citation peak in Australia, although this data was not field normalised (Gu & Blackmore, 2017). Since a previous study has shown different career citation patterns in some respects for different disciplines in the USA (Kolesnikov, Fukumoto, & Bozeman, 2018), the overall decreasing trend found here may well not apply to all US fields.

There are multiple possible explanations for the trends in the graphs and the apparently counterintuitive finding that average research impact per article decreases over careers, including the following.

- **Citation impact influence on career**: Producing low impact research, perhaps by accident, seems likely to directly influence the probability of an academic ceasing publishing. This could be due to failing tenure, failing to get a new job, moving to a teaching-oriented role or taking on administration or managerial roles.
- **Career influence on citation impact**: Low impact articles might be more applied in nature and therefore less cited (e.g., Moed, 2010), with the application signalling willingness to leave academia for a preferred outside career or spin-off company. Speculatively, senior researchers might (a) pursue mature research areas that are less cited, (b) mentor less capable PhD students, producing lower impact co-authored work, (c) devote less effort to their publishing after achieving career goals, or (d) author higher risk research with more chance of creating a highly cited paper even though most papers attract few citations.
- **Networking effects:** Assuming that a researcher ceasing publishing retires or moves to a non-academic job, their last publication may be less promoted by them to colleagues or friends.
- **Technical factors**: The last publication written by a researcher is the least likely to be self-cited (co-authors may self-cite), losing a source of citations. Given that papers typically have at least four authors, this seems unlikely to be a major factor. Nevertheless, reduced self-citations seem likely to make some decrease in average citation impact at the end of careers. Follow-up studies using data without self-citations would be useful to test this hypothesis. Moreover, if Scopus IDs tend to detach the early parts of careers (PhD, postdoctoral positions) from later positions, for example due to institutional moves, then the earliest publications of longer-term researchers may have lower impact than suggested by the data.

# 5 Conclusions

The career-long decline in average citation impact per article for the US overall is a key new finding, although it may be partly due to fewer self-citations to a researcher's last output (but co-authors may self-cite it). This effect for short-term or medium-term researchers is not concerning, since they leave academic publishing, perhaps because they struggle to produce higher impact work. The long-term researcher results are more important, however. Of course, the declining average citation impact per article finding is a statistical average phenomenon and a substantial minority of US academics will follow more positive career trajectories.

The tendency for average citation impact per article to decline over careers for academics with first and last journal articles published from the USA is a potential issue for academic decision makers. Since journal article publishing may be minor part of the role of senior academics, this is not a personnel concern. Nevertheless, it suggests that policy focusing on creating high impact work should consider prioritising junior and perhaps mid-career academics. This adds new support to concerns previously raised about the domination of biomedical funding by senior academics (Levitt & Levitt, 2017), for example. It also confirms the need for funding programmes targeting early career researchers, such as one from the Department of Energy (science.osti.gov/early-career).

As apparently the first study with this analysis approach (all researchers in a country, tracking cohorts from first to last publication, separately by year), it is important to follow up the results by identifying national and disciplinary differences. Most importantly, the major causes of the fall in average citation impact need to be identified so that the results can be properly interpreted in context.

**DATA AVAILABILITY**
The processed data used to produce the tables and figures are available in the supplementary material (https://doi.org/10.6084/m9.figshare.11791059). A subscription to Scopus is required to replicate the research, except with updated citation counts, with the methods described above.

16